\documentclass[aps,twocolumn,showpacs]{revtex4}
\usepackage{graphicx,epsfig,amsmath,amsfonts,epic,eepic,psfrag,latexsym,amssymb}

\begin{document}

\title{Dynamics of the BCS-BEC crossover in a degenerate Fermi gas}

\author{M. H. Szyma{\'n}ska${}^{1}$}
\author{B. D. Simons${}^{1}$}
\author{K. Burnett${}^{2}$}

\affiliation{${}^{1}$ Cavendish Laboratory, University of Cambridge,
Madingley Road, Cambridge CB3 OHE, UK \\ 
${}^{2}$ Clarendon Laboratory, University of
Oxford, Parks Road, Oxford, OX1 3PU, UK}

\begin{abstract}
We study the short-time dynamics of a degenerate Fermi gas positioned
near a Feshbach resonance following an abrupt jump in the atomic
interaction resulting from a change of external magnetic field. We
investigate the dynamics of the condensate order parameter and pair
wavefunction for a range of field strengths. When the abrupt jump is
sufficient to span the BCS to BEC crossover, we show that the rigidity
of the momentum distribution precludes any atom-molecule oscillations
in the entrance channel dominated resonances observed in the
${}^{40}$K and ${}^{6}$Li. Focusing on material parameters tailored to
the ${}^{40}$K Feshbach resonance system at $202.1$\, gauss, we
comment on the integrity of the fast sweet projection technique as a
vehicle to explore the condensed phase in the crossover region.
\end{abstract}

\maketitle

Ultracold alkali atomic gases provide a valuable arena in which to
explore molecular Bose-Einstein condensation (BEC)~\cite{Greiner03}
and fermionic pair condensation~\cite{Regal04-2,Zwierlein04}. Further,
the facility to control interparticle interactions via a
magnetically-tuned Feshbach resonance (FR) provide a unique
opportunity to investigate the BCS-BEC crossover and the dynamics of
condensate formation. As well as the adiabatic association of
molecules~\cite{Greiner03}, both fast sweep `projections' of fermionic
pair condensates onto the molecular BEC~\cite{Regal04-2,Zwierlein04},
and atom-molecule Ramsey fringes~\cite{Donley} have been reported in
the recent literature. Lately, motivated by earlier work on the
mean-field BCS system~\cite{Levitov1}, it was shown in separate
works~\cite{Levitov2,Anton} that the mean-field equations of motion of
a Bose-Fermi (BF) model, commonly used to describe the FR system, are
characterised by an integrable nonlinear dynamics. From these works,
three striking predictions emerged: Firstly, when perturbed by an
abrupt change in the strength of the pair interaction, the condensate
order parameter exhibits substantial oscillations which range in
magnitude between some initial state value, $\Delta_\mathrm{I}$, and
that expected for the equilibrium final state configuration
$\Delta_{\mathrm{eq}}$. Secondly, in the absence of energy relaxation
processes, these oscillations remain undamped suggesting the potential
to observe coherent atom-molecule oscillations in the FR system.
Thirdly, it is proposed that a spectral `hole-burning' phenomena in
the atomic momentum distribution, at a frequency associated with one
half of the molecular binding energy, provides a signature of such
atom-molecule oscillations~\cite{Anton}.

In the following, we will argue that this behaviour rests on an
auxiliary constraint that, in the present system, seems hard to
justify. Drawing on the results of a numerical analysis of the
unconstrained dynamics, we will show that, in the absence of
relaxational processes, the oscillations of the order parameter are
damped substantially, even at the level of the mean-field.  When the
abrupt change of the interaction is sufficient to span the BCS-BEC
crossover, for both the single channel and BF systems, the magnitude
of oscillations remain small attenuating to some value $\Delta_{\rm
F}$ which lies close to the initial state, and much smaller than
the expected equilibrium state value, $\Delta_{\rm eq}$, while the
distribution remains essentially ``frozen'' to the initial state.  The
hole-burning oscillations of the form predicted in Ref.~\cite{Anton}
do not appear in either regime. From these results, we are able to
conclude that, although the observation of BCS-BEC like atom-molecule
oscillations in alkali Fermi gases seems infeasible, the rigidity of
the initial state distributions validate fast sweep
techniques~\cite{Regal04-2} as a reliable experimental tool to explore
the crossover region.

Formally, a detailed microscopic theory of FR phenomena demands
consideration of all matrix elements connecting different spin states
participating in condensate formation. In practice, an accurate
description of the resonance can be obtained either by using a
single-channel magnetic field dependent effective interaction between
atoms in the entrance channel or, more widely, in the two-most
relevant channels. For low relative momenta, relevant to cold atom
physics, the full form of complex atomic potentials are not
resolved. Indeed, separable potentials with microscopic parameters
drawn from experiment and exact multichannel calculations can be used
to recover all low-energy binary scattering
observables~\cite{Oxford1,Oxford2}. Since, very often, only one bound
state of the closed channel potential is relevant, it has been
traditionally replaced by a fictitious Bose-particle and FR phenomena
captured by an effective two-channel BF
Hamiltonian~\cite{Holland,meera}.

Now, applied to the entrance channel dominated resonances observed in
the ${}^{40}$K and ${}^{6}$Li system, formal
calculations~\cite{Oxford1,Oxford2} support a picture in which the
BCS-BEC crossover \emph{is mediated by only a very small admixture of
closed channel states} --- e.g., in the ${}^{40}$K FR at 202.1 gauss,
the admixture of the closed channel is ca.~$8\%$ or less of the total
\cite{Oxford2}. Rather, the weakly bound molecular state appears at a
detuning $E_0$ which lies far from the value where the resonance state
of the closed channel crosses the dissociation threshold. Since the
two-body observables drawn from the exact numerical solution of the
Shr\"odinger equation within finite-range single and two-channel
models do not differ over a wide range of fields~\cite{Oxford2}, FR
phenomena in ${}^{40}$K can be equally-well described by a
single-channel theory,
\begin{equation}
\hat{H}=\sum_{\mathbf{k}s} \epsilon_{\mathbf{k}} 
a^{\dagger}_{\mathbf{k}s} a_{\mathbf{k}s}
+\sum_{\bf{k}\bf{k'}\bf{q}} V_{{\bf{k}}{\bf{k}}'} 
a^{\dagger}_{{\bf{k}}+{\bf{q}}\uparrow} 
a^{\dagger}_{-\bf{k}\downarrow} a_{-{\bf{k}}'\downarrow} 
a_{{\bf{k}}'+{\bf{q}}\uparrow},
\label{HMB1ch}
\end{equation}
involving Fermi operators $a^{\dagger}_{\mathbf{k}s}$ and
$a_{\mathbf{k}s}$.  In the following, to account for the entire region
of crossover (and not only the universal regime) we take as matrix
elements $V_{{\bf{k}}{\bf{k}}'}=V_0(B)\chi_{\bf
k}(\sigma_{\mathrm{bg}}) \chi_{{\bf k}'}(\sigma_{\mathrm{bg}})$ with
$\chi_{\bf k}( \sigma_{\mathrm{bg}})=\exp[-({\bf
k}\sigma_{\mathrm{bg}})^2/2]$ and the parameters $V_0(B)$ and
$\sigma_{\mathrm{bg}}$ chosen to recover the correct magnetic field
dependence of the scattering length and the highest vibrational bound
state~\cite{Oxford2}.

Applied to the fields $\Phi_{\bf k}=\sum_s\langle a^{\dagger}_{\bf{k}
s}a_{\bf{k}s}\rangle$/2 and $\kappa_{\bf{k}}=\langle a_{-\bf{k}
\downarrow}a_{\bf{k},\uparrow}\rangle$, the Heisenberg equations of motion
$i\langle\dot{A}\rangle=\langle[A,\hat{H}]\rangle$ translate to the 
relations
\begin{eqnarray}
i\dot{\kappa}_{\bf{k}} &=& 2\epsilon_{\bf{k}}\kappa_{\bf{k}} -
\Delta_{\bf{k}}(2\Phi_{\bf{k}}-1)\nonumber \\ i\dot{\Phi}_{\bf{k}} &=&
\Delta_{\bf{k}} \kappa_{\bf{k}}^{\star} -
\Delta^{\star}_{\bf{k}}\kappa_{\bf{k}},
\label{eqofmotion}
\end{eqnarray}
where $\Delta_{\bf k}=V_0(B)\chi_{\bf
k}(\sigma_{\mathrm{bg}})\sum_{{\bf k}'} \chi_{{\bf
k}'}(\sigma_{\mathrm{bg}}) \kappa_{{\bf k}'}$ denotes the complex
order parameter.  Similarly for a BF theory, defining $g_{\bf
k}=g_0\chi_k(\sigma)$ as the coupling of the entrance channel states
to the bosonic field $b_{\bf k}$  associated with the Feshbach
resonance level configuration of the closed channel, the equations of
motion acquire the same form as~(\ref{eqofmotion}) with $\Delta_{\bf
k} = g_0 \chi_{\bf k}(\sigma) b_0 +V_{\mathrm{bg}}\chi_{\bf
k}(\sigma_{\mathrm{bg}})\sum_{{\bf k}'} \chi_{{\bf
k}'}(\sigma_{\mathrm{bg}}) \kappa_{{\bf k}'}$ and $b_0=\langle b_{{\bf
k}=0}\rangle$ obeying the supplementary equation $i\dot{b}_0=E_0(B)
b_0 +g_0\sum_{\bf{k}} \chi_{\bf k}(\sigma) \kappa_{\bf k}$. As with
the single-channel theory, the five parameters which characterise the
resonance, the background potential strength $V_{\mathrm{bg}}$ and
range $\sigma_{\mathrm{bg}}$ (which define the entrance channel
scattering length and its highest vibrational bound state), the
interchannel coupling $g_0$ and its range $\sigma$, and the detuning
$E_0(B)$ (which specifies the position and width of the resonance),
are determined from experiment and exact multichannel
calculations~\cite{Oxford2}.

Before turning to the numerical analysis, it is instructive to
contrast the present approach to that adopted in
Refs.~\cite{Levitov1,Levitov2,Anton}.  Given an initial condition, the
Heisenberg equations of motion~(\ref{eqofmotion}) present a
deterministic time evolution of the density distributions. Indeed, the
conservation of total atomic density $n$, implicit in the
dynamics~(\ref{eqofmotion}), provides a check on the integrity of the
numerical integration described below. By contrast, the integrability
of the equations of motion as described by
Refs.~\cite{Levitov1,Levitov2,Anton} relies on an additional
constraint involving the density and an auxiliary parameter playing
the role of a ``chemical potential''.  The need for this constraint
originates from the fact that, as with the equilibrium steady state
solution, the analytical solution of the dynamical equations is
determined only up to an arbitrary momentum dependent sign factor
which is fixed by the auxiliary constraint \cite{Anton}, instead of
naturally following from the initial conditions. A similar
phenomenology describes the effect of a classical laser source on a
semiconductor electron/hole system where the laser frequency
``imprints'' a chemical potential onto the system~\cite{SchR}. While
equilibration processes are sufficiently small, the electrons and
holes assume a non-equilibrium distribution around the externally
imposed chemical potential resulting in a phenomenon of
``spectral-hole burning'' in which the density distribution is
depleted at the laser frequency. In the atomic gas system, where the
degrees of freedom remain internal, it is difficult to see how such a
choice is motivated or justified.  Crucially, we will see that the
unconstrained dynamics associated with the equations of
motion~(\ref{eqofmotion}) lead to physical behaviour very different
from that obtained from the constrained
dynamics~\cite{Levitov1,Levitov2,Anton}.

With this background, let us now turn to the results of the numerical
investigation of the dynamics~(\ref{eqofmotion}) performed using an
adaptive step Runge-Kutta algorithm applied on a grid in momentum
space chosen fine enough to ensure convergence at each time step.
Although we find that the qualitative behaviour of BCS and BF dynamics
is generic, we focus specifically on potentials characteristic for the
${}^{40}$K resonance at $B_0=202.1\,$gauss with a density of
$n=1.5\times 10^{13}\,{\rm cm}^{-3}$ (i.e. $T_{\mathrm{F}}=0.35\,\mu
\mathrm{K}$) comparable to that used in
experiment~\cite{Regal04-2}. With these parameters, the equilibrium
properties of the effective single-channel and BF model essentially
coincide (for further details and values of parameters, we refer to
Ref.~\cite{Oxford2}). Therefore, to keep our discussion concise, we
will focus on the single-channel theory noting that the parallel
application to the BF model with appropriate physical parameters
generates \emph{quantitatively} similar results.
\begin{figure}[htbp]
  \centering
  \includegraphics[width=3in]{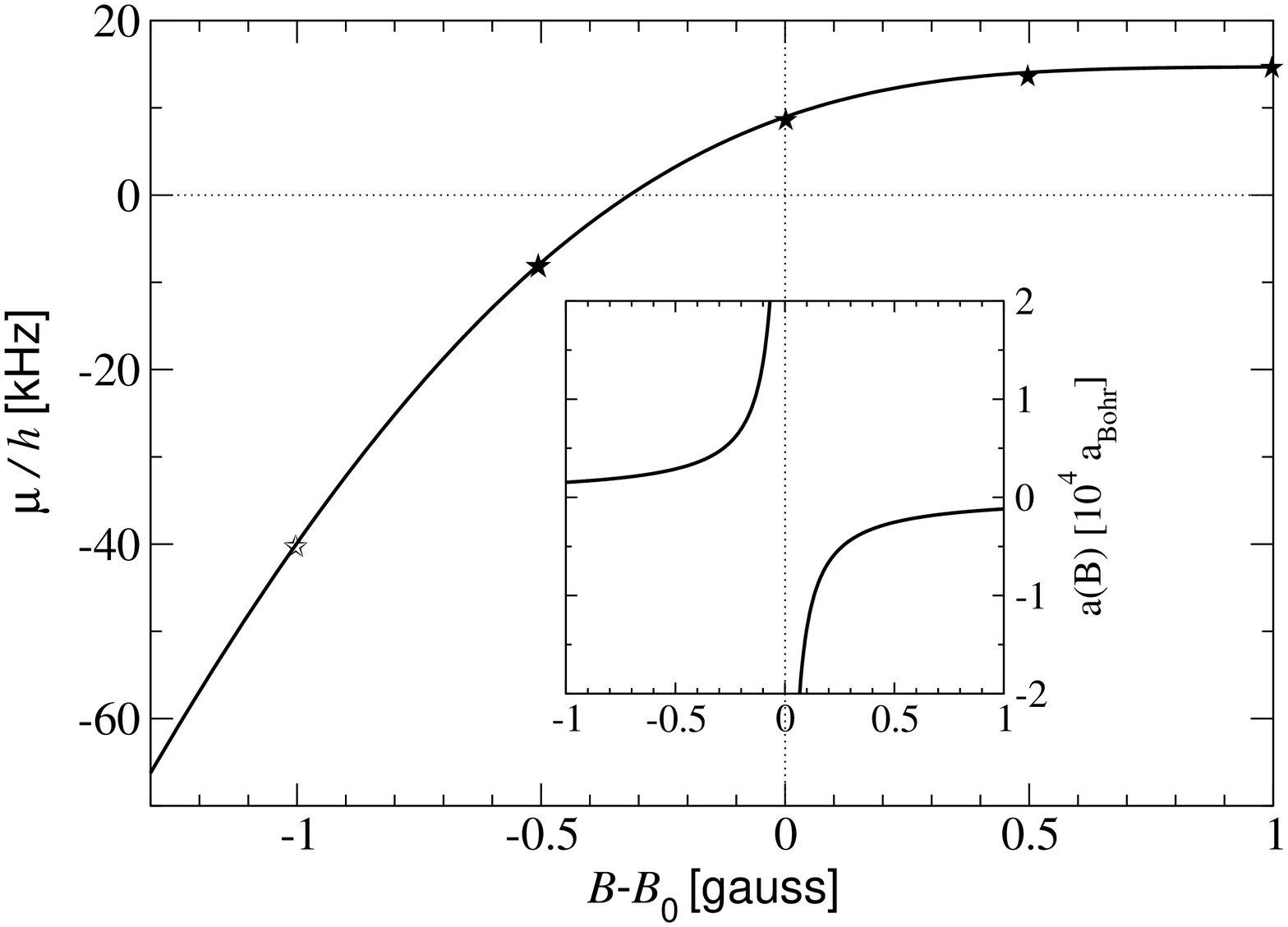}
  \caption{Variation of the chemical potential $\mu$ with field $B$ at
  $T=0$ for a mixture of fermionic $^{40}$K atoms prepared in the
  $(f=9/2,m_f=-9/2)$ and $(f=9/2,m_f=-7/2)$ Zeeman states at a density
  of $n=1.5\times 10^{13}\,{\rm cm}^{-3}$
  (i.e. $T_{\mathrm{F}}=0.35\,\mu \mathrm{K}$). The FR takes place at
  a field $B_0=202.1\,$gauss while the BEC-BCS crossover ($\mu=0$)
  occurs when $B-B_0\simeq 0.3\,$gauss. The inset indicates the
  scattering length $a(B)=a_{\mathrm{bg}}(1-\frac{\Delta B}{B-B_0})$
  in the universal regime, where a$_\mathrm{bg}$ is the background
  potential scattering length and $\Delta$B the width of the
  resonance. Note that, in the present theory, we use finite range
  potentials with parameters which describe the FR also far outside
  the universal region \cite{Oxford2}.}
  \label{fig:chem}
\end{figure}
At a field of ca.~$1.0\,$gauss above the FR, the condensate has an essentially 
BCS-like character while the experiment using the fast sweep technique 
observed the condensate starting from $0.5\,$gauss above $B_0$. To explore 
the entire region of interest, we choose as initial conditions field values 
$B_\mathrm{I}$ which span the entire crossover region (marked by stars 
in Fig.~\ref{fig:chem}). Starting from the ground state $T=0$ distribution, 
we follow the dynamics of the condensate after an abrupt switch to some
different value of magnetic field $B_\mathrm{F}$.

\begin{figure}[htbp]
  \centering
  \includegraphics[width=3.2in]{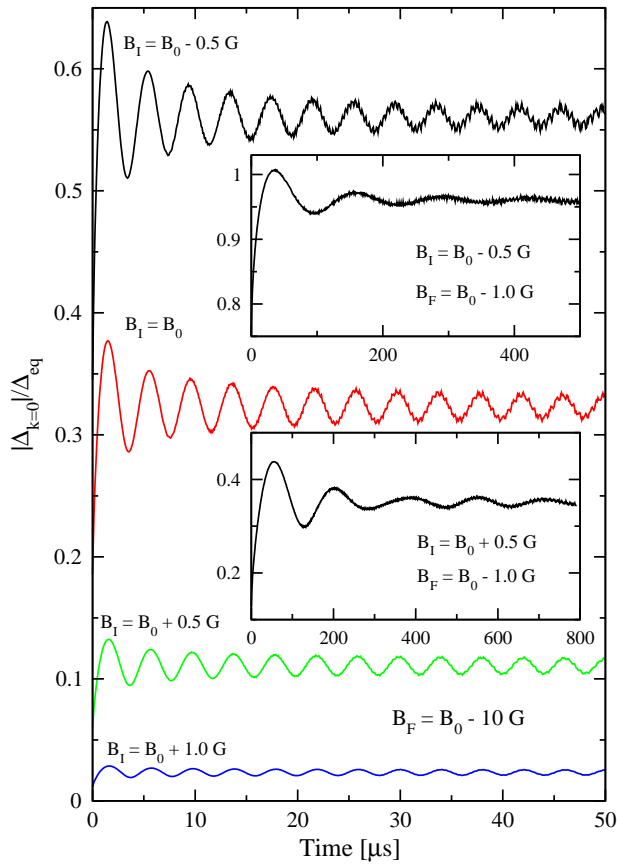}
  \caption{Time dependence of the order parameter amplitude
    $|\Delta_{{\bf k}=0}|/\Delta_{\rm eq}$ following an abrupt switch
    from a field of $(B_\mathrm{I}-B_0)/{\rm gauss}=-0.5$ (top),
    $0.0$, $0.5$, $1.0$ (bottom) to $(B_\mathrm{F}-B_0)/{\rm
    gauss}=-10$, and from $(B_\mathrm{I}-B_0)/ {\rm gauss}=-0.5$ (top
    inset) and $(B_\mathrm{I}-B_0)/{\rm gauss}=0.5$ (bottom inset) to
    $(B_\mathrm{F}-B_0)/{\rm gauss}=-1.0$. In addition to the constant
    phase velocity of $\Delta_{{\bf k}=0}$ (found numerically to be
    set by the final state equilibrium chemical potential $\mu$),
    there is an additional time-dependent phase modulation whose
    characteristics mirror closely that of the amplitude oscillations. }
  \label{fig:delta}
\end{figure}

\begin{figure}[htbp]
  \centering
  \includegraphics[width=3.2in]{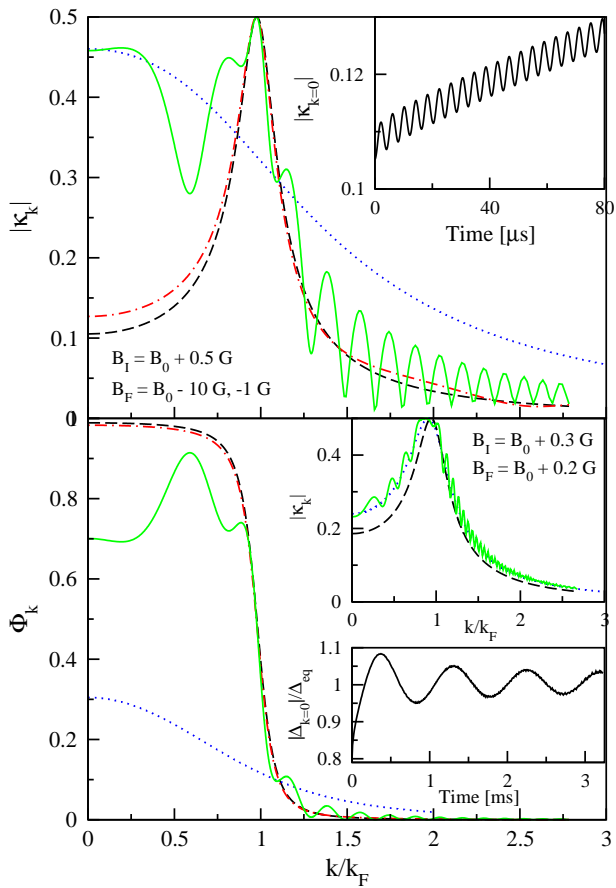}
  \caption{The pair wave-function $|\kappa_{\bf{k}}|$ (upper panel)
  and the density distribution function $\Phi_{\bf k}$ (lower panel)
  shown as function of $k=|{\bf k}|$ with $B_\mathrm{I}-B_0=0.5$\,
  gauss (dashed lines) and $B_\mathrm{F}-B_0=-10\,$ gauss after
  $50\,\mu$s (dashed-dotted line) and $B_\mathrm{F}-B_0=-1.0$\, gauss
  after $800\,\mu$s (solid line) following the abrupt switch as in
  Fig.  \ref{fig:delta}. The dotted lines signify the ground state
  equilibrium distributions at $B_\mathrm{F}-B_0=-1.0$\, gauss
  included for comparison.  The upper inset shows oscillations of
  $|\kappa_{{\bf k}=0}|$ for $B_\mathrm{F}-B_0=-10$\, gauss. The lower
  two insets refer to a rather weak perturbation on the BCS side from
  $B_\mathrm{I}-B_0=0.3$\, gauss (dashed line) to
  $B_\mathrm{F}-B_0=0.2$\, gauss. The dotted lines shows the
  equilibrium distribution while the solid line provides a snap-shot
  of the distribution at $3$\,ms after the switch. Note that the
  harmonic modulations visible in the distribution functions translate
  to a single energy scale of the same order of magnitude as the
  period of oscillations seen in $|\Delta_{{\bf k}=0}|$. }
  \label{fig:ukvk}
\end{figure}

Figure~\ref{fig:delta} shows the time-evolution of the order parameter
$|\Delta_{{\bf k}=0}|$, normalised by the value,
$\Delta_{\mathrm{eq}}$ (i.e. the value that it would acquire were the
system to reach the $T=0$ ground state at the final field
$B_\mathrm{F}$).  Here we have chosen a field
$B_\mathrm{F}-B_0=-10\,$gauss deep within the BEC phase where the
large binding energy of molecules allows their momentum distribution
to be inferred from time of flight measurements~\cite{Regal04-2}. In
contrast to the predictions of the constrained
dynamics~\cite{Levitov1,Levitov2,Anton}, these results show that (a)
the coherent oscillations are substantially damped even at the level
of mean-field, (b) the amplitude of the oscillations is small, and (c)
the order parameter $|\Delta_{{\bf k}=0}|$ asymptotes to a value much
less than the expected final state equilibrium value
$\Delta_{\mathrm{eq}}$. Referring to the insets of
Fig.~\ref{fig:delta}, one may note that, when the initial and final
conditions are drawn closer, the period of the oscillations becomes
longer and the effects of the damping more pronounced. Moreover,
although the period of the oscillations increases monotonically with
$\Delta_{\rm eq}$, the dependence is nonlinear and, referring to the
bottom inset of Fig.~\ref{fig:ukvk}, one may note that
$\Delta_{\mathrm{eq}}$ does not provide a ceiling for the magnitude of
the oscillation.

To interpret the generic behaviour, it is instructive to access the 
time-dependence of the pair wavefunction $\kappa_{\bf k}$ and the 
distribution function $\Phi_{\bf k}$.
Figure~\ref{fig:ukvk} shows that, when the abrupt switch takes place
from $B_{\mathrm{I}}-B_0=0.5\,$gauss to
$B_{\mathrm{F}}-B_0=-10\,$gauss, although there is a slight tendency
to shift towards the final state equilibrium distribution, the pair
wavefunction and the density distribution remain essentially frozen
close to the initial BCS-like distribution, exhibiting only small
oscillations in time. In the absence of energy relaxational processes,
the system is unable to significantly redistribute weight.  By
contrast, when the abrupt switch takes place from
$B_{\mathrm{I}}-B_0=0.5\,$gauss to $B_{\mathrm{F}}-B_0=-1.0\,$gauss,
the proximity of the two phases allows the system to converge to a
(non-stationary) modulated distribution whose (stationary) envelope
reflects more closely the final state equilibrium distribution.
Referring to insets in the bottom panel one may note that a weakly
perturbed condensate on the BCS side attenuates to its equilibrium
value with the pair distribution showing only small modulations around
the equilibrium in accord with the linear stability analysis of the
weakly perturbed BCS system~(\ref{eqofmotion}) discussed in
Ref.~\cite{Volkov}.  In particular, one may note that here (and,
indeed, for other values of the initial and final conditions) the
hole-burning phenomenon predicted by the constrained
dynamics~\cite{Anton} does not appear.

To assess the potential to observe coherent atom-molecule oscillations, one
can monitor the time evolution of the number of condensed molecules,
\begin{equation}
n_{\mathrm{mc}}(t)=\left|\int d^3 k \ \kappa_k(t)\,
\phi_\mathrm{B}(k,B_\mathrm{F})\right|^2\,.
\label{nmc}
\end{equation}
Here $\phi_\mathrm{B}(k,B_\mathrm{F})$ denotes the wavefunction of 
the highest vibrational bound state being an exact eigenstate of the 
two-body problem~\cite{Oxford2}. 
\begin{figure}[htbp]
  \centering
  \includegraphics[width=3.2in]{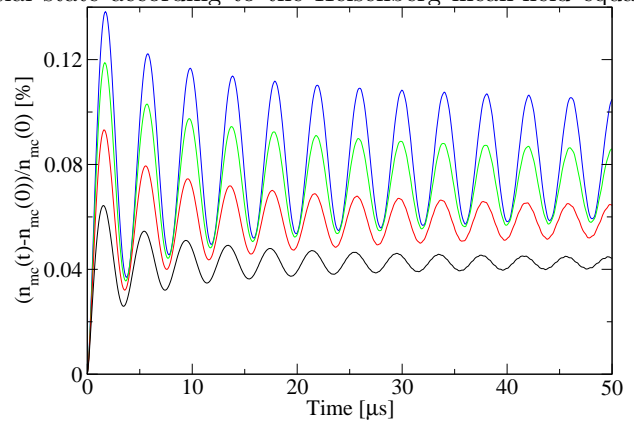}
  \caption{Time-dependence of the relative number of condensed
  molecules
  $(n_\mathrm{mc}(t)-n_\mathrm{mc}(0))/n_\mathrm{mc}(0)$. Here we have
  used the same field values as that used in Fig.~\ref{fig:delta}
  (main) with $(B_{\rm I}-B_0)/{\rm gauss}=-0.5\,$ (top), $0.0$, $0.5$,
  and $1.0$ (bottom) and $(B_{\rm F}-B_0)/{\rm gauss}=-10$. For these
  field values, when normalised to one half of the total atomic
  density, $n_\mathrm{mc}(0)=0.0005$, $0.012$, $0.1$, and $0.3$
  respectively.}
  \label{fig:mol}
\end{figure}
Referring to Fig.~\ref{fig:mol}, one may note that the amplitude of
oscillations is negligible (less than one percent).  Although one may
adjust the final field $B_{\rm F}$ to lie closer to the FR, the
amplitude of the oscillations increases only slightly while the
damping rate is also enhanced. Moreover, the inclusion of processes
beyond the mean-field approximation considered here would simply
increase the damping rate and not enhance the oscillations. One may
therefore conclude that the observation of atom-molecule oscillations,
following an abrupt change in the interaction strength, is infeasible
for the entrance channel dominated resonances currently discussed in
the context of the ${}^{40}$K and ${}^{6}$Li system.

To conclude, we have presented a numerical analysis of the dynamical
mean-field equations for the single-channel theory of the FR following
an abrupt field change. In the range of physical parameters
appropriate to the ${}^{40}$K system, the consideration of the
two-channel BF (Bose-Fermi) theory does not change the results
\emph{quant}itatively. When applied to a theoretical regime where the
population of the closed channel states below resonance is high, the
numerical findings do not change \emph{qual}itatively. Relying on the
deterministic time-evolution of the initial state according to the
Heisenberg mean-field equations of motion, our results differ
substantially from the findings of the constrained dynamics considered
previously~\cite{Levitov1,Levitov2,Anton}. Over a wide range of
initial conditions, we observe substantially damped oscillations with
an amplitude strongly dependent on initial conditions and a frequency
set by the interaction strength after the switch. To assess the
capacity for BCS-BEC like atom-molecule oscillations following an
abrupt change in the interaction, we have chosen initial and final
conditions to span the crossover from the BCS to the BEC limits. We
have found that the amplitude of atom-molecule oscillations is
negligible and the distribution is essentially frozen to the initial,
a behaviour reminiscent of an orthogonality catastrophe.  We conclude
that, in the entrance channel dominated resonances observed in
${}^{40}$K and ${}^{6}$Li, the observation of atom-molecule
oscillations is infeasible. The rigidity of the condensate
wavefunction distribution at short-time scales, where the time
evolution is mean-field in character, supports the method of the fast
sweep as a reliable technique to probe the fermionic pair condensate.

We are grateful to Krzysztof G{\'o}ral and Thorsten K\"ohler for
numerous discussions concerning the FR physics, to Peter Littlewood
for stimulating discussions and to S{\l}awomir Matyja{\'s}kiewicz for
advice concerning the numerical techniques. This research has been
supported by Gonville and Caius College Cambridge (M.H.S.)


\begin{thebibliography}{20}
\bibitem{meera} Note that the particular architecture of the hyperfine 
states in the $^{40}$K system calls into question the validity of a 
BF model as a microscopic theory~\cite{meera2}.
\end{thebibliography}
\end{document}